# Об одной интересной и важной модели в квантовой механике. I. Динамическое описание

Юрий Григорьевич Рудой, Енок Олуволе Оладимеджи (мл.)

Российский университет дружбы народов (РУДН)
117198 Россия, Москва, ул. Миклухо-Маклая, 6; e-mail: rudikar@mail.ru, nockjnr@gmail.com

В данной статье подробно рассмотрена одна из наиболее интересных моделей в нерелятивистской квантовой механике одной массивной частицы (в одномерном варианте), а именно модель, введенная Г. Пёшлем и Э. Теллером в 1933 году. Эта модель обладает рядом интересных свойств – в частности, ее предельными случаями являются две наиболее известные модели, изучаемые во всех (в том числе начальных) курсах квантовой механики, в том числе и для ряда инженерных и технических специальностей (в том числе по нанотехнологиям). Этими моделями являются, во-первых, сохраняющая конфайнмент модель квазисвободной частицы в «ящике» с непроницаемыми стенками и, во-вторых, модель квантового гармонического осциллятора, введенная Ф. Блохом в 1932 году и не обладающая конфайнментом. В статье предложено элементарное, но детальное рассмотрение взаимосвязей потенциала, волновых функций и энергетического спектра всех указанных моделей. Одним из основных результатов является введение в моделях с конфайнментом оператора давления на основе теоремы Гельмана–Фейнмана и анализ предельных случаев для этой величины. В следующей работе предполагается обобщение полученных результатов на случай отличной от нуля температуры.

*Ключевые слова*: гармонический квантовый осциллятор Блоха, ангармонический квантовый осциллятор, квантовый осциллятор Пёшля–Теллера, оператор давления, теорема Гельмана–Фейнмана.

## Введение

В связи с интенсивным развитием нанотехнологий (см., например, [1]), в настоящее время вновь повысился интерес к описанию квантовых объектов, состоящих из одной или нескольких микрочастиц и находящихся во внешнем силовом потенциале. Если ранее подобные задачи встречались в основном в физике атомного ядра и элементарных частиц, то сейчас они стали характерны для проблем физики конденсированного состояния, а также для объектов, находящихся в искусственных квантовых ловушках.

Типичным примером могут служить *квантовые точки* в гетероструктурах [1], причем подобные квантовые объекты являются, как правило, низкоразмерными $D < 3$ (ниже мы ограничимся одномерным случаем $D = 1$). Для указанного класса задач представляет интерес, во-первых, наличие т.н. *управляющих параметров* – например, массы частицы $m$ и интенсивность потенциала $V_0$, и, во-вторых, возможность «термодинамической» постановки всей задачи в целом. Последнее предпола-гает пространственную ограниченность, или *конфайнмент*, квантового объекта в области с характерной шириной $L$, а также возможность помещения указанного квантового объекта в термостат с температурой $T$ (по шкале Кельвина). Существенно, что лишь для ограниченных объектов (с конечным $L$) имеется возможность введения понятия *механического давления P*, что делает «термодинамическую» постановку задачи более последовательной – в частности, позволяет проводить замкнутые термодинамические циклы (примеры подобных механических циклов – Отто и Джоуля–Брайтона рассмотрены в следующей статье).

Наиболее известными (см., например, [2–4]) – и притом точно решаемыми (!) – задачами одномерной квантовой механики, являются следующие две задачи: 1) задача о (квази)свободной частице в «ящике» шириной $L$, а также 2) задача о гармоническом квантовом осцилляторе с собственной частотой $\omega_0$, определенная на всей оси ($L \to \infty$) – т.н. «осциллятор Блоха», введенный Ф. Блохом в 1932 году.

Очевидно, что лишь первая из задач обладает конфайнментом (и, следовательно, давлением), который обеспечивается непроницаемыми для частицы стенками «ящика», что формально соответствует т.н. *дельтаобразному* потенциалу, бесконечно высокому на стенках и равному нулю внутри ящика; разумеется, подобный потенциал носит несколько искусственный характер, но допускает приближенный термодинамический анализ. Что касается второй задачи, то ее решение в «термодинамическом» варианте при ненулевой температуре $T$ является одним из немногих (если не единственным) примером *точного* квантово-статистического расчета (см., например, [5], §37).

В связи с изложенным, на наш взгляд, представляет несомненный интерес рассмотрение еще одной одномерной квантово-механической задачи с потенциалом, предложенным Пёшлем и Теллером в 1933 году, который объединяет две указанные выше задачи и при этом допускает точное (!) решение, описанное в двух широко известных учебных пособиях по квантовой механике [2–4]. Заметим, что с математической точки зрения задача об осцилляторе Пёшля–Теллера не намного сложнее по сравнению, например, с задачей об осцилляторе Блоха, однако с физической точки зрения указанная задача является значительно более общей и содержательной.

## Описание квантовой модели Пёшля–Теллера и ее предельных случаев

В нерелятивистской квантовой механике в одномерном случае волновая функция $\psi(x)$ и энергетический спектр частицы $E_n$ описывается стационарным уравнением Шрёдингера с потенциалом $V(x)$. Если потенциал $V(x)$ неограничен сверху и стремится к *бесконечным* значениям на концах области определения – например, на всей оси $x$ для осциллятора Блоха (БО) или на каком-либо отрезке с конечной шириной $L$ для

«свободной» частицы в ящике (СЧ), то волновая функция ψ(x) обращается в нуль при $x \to \pm\infty$ (для случая БО) или при $x_\pm(L) = \pm 1/2 L$ (для случая СЧ).

Благодаря этому спектр $E_n$ является чисто *дискретным* и, как и потенциал $V(x)$, также неограниченным сверху, так что главное квантовое число $n$ принимает все значения от 0 (или 1) до ∞. Именно подобная ситуация в целом характерна для модели сильно нелинейного (по существу, даже сингулярного) осциллятора Пёшля–Теллера (ПТО), потенциал которого имеет следующий вид, где *эффективная ширина L* имеет смысл управляющего параметра:

$$V_{\text{ПТ}}(x; L) = V_0 \text{tg}^2[\alpha(L)x], \quad 0 \leq V_0 < \infty; \quad \alpha(L) = \pi/L;$$

$$x_-(L) \leq x \leq x_+(L), \quad x_\pm(L) = \pm 1/2 L, \quad 0 < L < \infty. \tag{1}$$

Потенциал $V(x; L)$ обладает рядом весьма простых и полезных свойств, которые существенно упрощают решение уравнения Шрёдингера для ПТ-осциллятора. Прежде всего, потенциал $V(x; L) = V(-x; L)$ является *симметричным*, или чётным, относительно точки $x = 0$, причём эта точка является единственным (и, следовательно, абсолютным) *минимумом*, в котором $V(0; L) = 0$ для всех L; кроме того, потенциал $V(x; L)$ обладает парой симметричных *сингулярностей* $V(x; L) \to \infty$ для значений $x_\pm(L)$, поскольку $\text{tg}(\pi/2) \to \infty$.

Наконец, потенциал $V(x; L)$ и все его производные $V^{(n)}(x; L)$ по $x$ порядка $n \geq 1$ являются *гладкими* для всех значений $x_-(L) < x < x_+(L)$ при всех отличных от нуля значениях *эффективной ширины L*, имеющего смысл управляющего параметра. Заметим, что зависимость величины $\alpha(L) = \pi/L, 0 < \alpha(L) < \infty$, непосредственно определяющей потенциал $V(x; L)$ из (1), имеет свойство убывания при возрастании ширины L, так что $d\alpha(L)/dL = -\alpha(L)/L < 0$ для всех L, причём $\alpha_{\min}(\infty) = 0$ при $L \to \infty$.

Покажем теперь, что модель ПТО при определённых условиях сводится к ранее упомянутым известным моделям, а именно к модели БО вблизи точки $x = 0$ для малых значений $x/L \ll 1$ и, соответственно, к модели СЧ в ящике вблизи точек $x_\pm(L)$, т.е. при малых значениях $|y|/L = |x|/L - 1/2 \ll 1$; рассмотрим эти два случая последовательно.

**1. Переход от осциллятора Пёшля–Теллера к осциллятору Блоха.** Учтём, что при фиксированном значении $\alpha(L) = \pi/L$ и при малых значениях $\alpha(L)x \ll 1$ имеет место разложение $\text{tg}[\alpha(L)x] \approx \alpha(L)x + 1/3[\alpha(L)x]^3 + \ldots$, так что вблизи минимума $x = 0$, т.е. для значений $|x|/L \ll 1$, потенциал $V_{\text{ПТ}}(x; L)$ принимает вид:

$$V_{\text{ПТ}}(x; L) \approx V_0[\alpha(L)x]^2\{1 + 2/3[\alpha(L)x]^2 + \ldots\}. \tag{2}$$

Очевидно, что потенциал (2) учитывает не только слагаемое вида $x^2$, обеспечивающее поведение типа гармонического осциллятора, но также и относительно малое слагаемое вида $x^4$, обеспечивающее поведение типа *ангармонического* осциллятора. Потенциал (2) является интересным объектом исследования, однако здесь важно, что в (2) предполагается *конечное* значение ширины $L$ (и, соответственно, $\alpha(L)$), тогда как осциллятор Блоха определен только на всей оси $x$; последнее означает переход к пределу $L \to \infty$ и, следовательно, $\alpha(L) \to 0$.

Однако если ограничиться только этим пределом, то из (2) следует, что $V(x;L) \to 0$ при всех $x$, что не представляет интереса в термодинамическом контексте, поскольку такой предел соответствует случаю полностью свободной частицы, которая описывается волной де-Бройля и непрерывным спектром энергии. Однако ситуация изменяется, если *одновременно* с пределом $L \to \infty$ и $\alpha(L) \to 0$ перейти к пределу $V_0 \to \infty$, так чтобы предел произведения $V_0 \alpha^2(L)$ был *конечен*. Положив его равным $1/2k$ = const, где $k$ – эффективный коэффициент упругости, нетрудно придти к решению для осциллятора Блоха, обладающего собственной частотой $\omega_0 = (k/m)^{1/2}$.

Таким образом, предельный переход от осциллятора Пёшля–Теллера к осциллятору Блоха описывается выражением для чисто параболического потенциала на всей оси $x$:

$$V_{\text{ПТ}}(x;L) \to V_{\text{ПТ}}(x;\infty) \equiv V_{\text{БО}}(x) = 1/2 kx^2 \quad (-\infty < x < \infty);$$

$$k = 2\pi^2(V_0/L)^2 = \text{const} \quad \text{при} \quad V_0, L \to \infty. \tag{3}$$

**2. Переход от осциллятора Пёшля–Теллера к квазисвободной частице в ящике.**
Как уже отмечалось выше, исходный потенциал (1) становится *сингулярным* в двух точках $x_{\pm}(L)$, что фактически означает переход к условиям обычного ящика с непроницаемыми стенками. С физической точки зрения модель Пёшля–Теллера предпочтительнее указанного «ящика», поскольку в ПТ-модели потенциал (1) сам «регулирует» поведение квантовой частицы и не нуждается в дополнительном введении весьма искусственных «стенок».

Как известно (см., например, [2–4]), в модели «ящика» внутри него вообще отсутствует какой-либо потенциал, а квантование энергии частицы всецело определяется физическими *граничными условиями*, согласно которым волновая функция $\psi(x)$ строго обращается в нуль в точках $x_{\pm}(L)$. Заметим, что квантование энергии – прежде всего, строгую положительность энергии $E_1(L)$ основного состояния частицы в ящике и дискретность всего спектра $E_n(L)$

$$E_1(L) \equiv W(L) = (\hbar^2/2m)\alpha^2(L), \quad E_n(L) = E_1(L)n^2 \quad (n = 1,2,...) \tag{4}$$

можно объяснить на основе соотношения неопределенностей в форме Гейзенберга (см., например, [2–4]).

Напомним, что *формально* потенциал (квази)свободной частицы в ящике может быть описан выражением

$$V_{\text{Сч}}(x;L) = 1/2\{\delta[x-x_+(L)] + \delta[x-x_-(L)]\} = \delta\{x^2 - [x_\pm(L)]^2\} = \delta[x^2 - 1/4L^2], \quad (5)$$
$$x_+(L) = -x_-(L) = 1/2L.$$

Обратим внимание на то, что в выражение (5) вообще не входит какая-либо «амплитуда» потенциала, что означает возможность вообще положить ее равной нулю $V_0 = 0$; действительно, внутри ящика это так (по определению модели), а на стенках сингулярность потенциала $V_{\text{Сч}}(x;L)$ столь сильна, что перекрывает любой предел $V_0 \to 0$.

С учетом этих соображений аппроксимируем исходный потенциал ПТ-модели (1) вблизи граничных точек $x_\pm(L)$, перейдя к новой переменной $y_\pm = [x - x_\pm(L)]$, которая, очевидно, мала вблизи точек $x_\pm(L)$. Учтем далее, что $\alpha(L)x = \alpha(L)[y + x_\pm(L)] = \alpha(L)y \pm \pi/2$, причем $\text{ctg}(\pi/2) = 0$, так что

$$\text{tg}[\alpha(L)x] = \{\text{tg}[\alpha(L)y \pm \pi/2]\} =$$
$$= \{\text{ctg}[\alpha(L)y] \pm [\text{ctg}(\pi/2)]\}\{\text{ctg}[\alpha(L)y]\text{ctg}(\pi/2) \pm 1\}^{-1} = \pm\text{ctg}[\alpha(L)y]. \quad (6)$$

Поскольку вблизи граничных точек величина $\alpha(L)y = \pi(y/L) \ll 1$, приближенное разложение для (6) имеет вид $\text{ctg}[\alpha(L)y] \approx [\alpha(L)y]^{-1} - 1/3[\alpha(L)y] + \ldots$. Учитывая здесь в качестве основного только первое (расходящееся) слагаемое, получаем для потенциала (1) ПТ-модели (вблизи граничных точек) следующее приближенное выражение:

$$V_{\text{ПТ}}(x;L) \to V_{\text{ПТ}}(x \to x_\pm(L);L) = V_{\text{Сч}}(y;L) \approx V_0(L/\pi)^2[1/(y_+)^2 + 1/(y_-)^2] =$$
$$= V_0(L/\pi)^2[(x_+(L)-x)^{-2} + (x_-(L)-x)^{-2}] =$$
$$= V_0(L/\pi)^2\{(x_+(L)-x)^2 + (x_+(L)+x)^2\}[(x_+(L)-x)^{-2}(x_+(L)+x)^{-2}]^{-1} = \quad (7)$$
$$= V_0(L^4/\pi^2)\{x^2 - [x_\pm(L)]^2\}^{-2}.$$

Завершая достаточно протяженный (хотя и вполне элементарный) расчет (7), получаем следующий результат:

$$V_{\text{ПТ}}(x;L) \to V_{\text{ПТ}}(x \to x_\pm(L);L) = V_{\text{Сч}}(x;L) \approx V_0(L^4/\pi^2)[x^2 - 1/4L^2]^{-2}; \quad (8)$$

сравнивая выражение (8) для потенциала частицы в ящике с его «точным» выражением (5), нетрудно видеть, что потенциал (8) обладает достаточно ярко выраженную сингулярность, по структуре совпадающую с (5), так что в пределе $|x| \to 1/2L$ можно *одновременно* принять $V_0 \to 0$ (сравните с аналогичным двойным пределом в выражении (3)).

Таким образом, содержание данного раздела можно кратко резюмировать следующим образом. Как видно из выражений (3) и (8), модель нелинейного сингулярного квантового осциллятора Пёшля–Теллера действительно *объединяет* две основных модели квантовой механики (в одном измерении), а именно: при *низких энергиях* (вблизи центра «эффективного ящика») модель ПТ-осциллятора воспроизводит модель линейного гармонического *осциллятора Блоха*, тогда как при *высоких энергиях* (вблизи стенок «эффективного ящика») модель ПТ-осциллятора воспроизводит модель (квази)свободной частицы в ящике. Как будет видно ниже, этот вывод подтверждается точным решением уравнения Шредингера для модели ПТ-осциллятора.

## Волновые функции и энергетический спектр квантовой модели Пёшля–Теллера

Уравнение Шредингера для одномерной модели ПТ-осциллятора имеет, как обычно, вид дифференциального уравнения 2-го порядка $(d^2/dx^2) + (2m/\hbar^2)[E - V_{\text{ПТ}}(x;L)]\psi(x) = 0$ с потенциалом $V_{\text{ПТ}}(x;L)$ из (1), точное решение которого можно найти в [2–4]. Наметим основные этапы и результат этого решения. Переходя к безразмерным переменным

$$\xi = \alpha(L)x = (\pi/L)x, \quad v(L) = V_0/W(L),$$
$$\varepsilon(L) = (E + V_0)/W(L) = [E/W(L)] + v(L), \qquad (9)$$

где $W(L)$ – энергия основного состояния (4) для частицы в ящике. Тогда с учетом равенства tg$^2$s = (1/cos$^2$s) − 1 уравнение Шредингера для волновой функции основного состояния $\psi_0$(s) принимает вид

$$d^2\psi/d\xi^2 + [\varepsilon(L) - v(L)/\cos^2\xi]\psi = 0. \qquad (10)$$

Для свободной частицы (при $V_0 = 0$, $v(L) = 0$) в ящике с конечной шириной $L$ решением (10) является $\psi_0(\xi) = \cos \xi$. Соответственно, при $v(L) \neq 0$ решение (10) должно иметь вид $\psi(\xi) = (\cos \xi)^{\lambda(L)}$, где показатель $\lambda(L) > 0$, поскольку согласно граничным условиям при $x = x_\pm(L)$ функция $\psi_0(\xi)$ должна обращаться в нуль при $\xi_\pm = \pm(\pi/2)$; из (10) видно, что показатель $\lambda(L) > 0$ (в частности, $\lambda(L) = 1$ при $v(L) = 0$) должен удовлетворять условию[1]

$$\lambda(L)[\lambda(L) - 1] = v(L), \quad \lambda(L) = 1/2\{1 + [1 + 4v(L)]^{1/2}\},$$
$$d\lambda(L)/dL = (1/L)\{2v(L)[1 + 4v(L)]^{-1/2}\}. \qquad (11)$$

Для общего случая возбужденных состояний можно показать [2–4], что полная волновая функция $\Psi_n(\xi)$ может быть представлена в виде произведения $\Psi_n(\xi) = C_n(\lambda(L))(\cos\xi)^{\lambda(L)} G_n^{[\lambda(L)]}(\sin\xi)$, где $G_n$ – полиномы Гегенбауэра, $C_n$ – нормировочные постоянные, однако для дальнейших термодинамических вычислений существенно знание только энергетического спектра $E_n^{\text{ПТ}}(L)$, который является чисто дискретным и неограниченным сверху ($n = 0, 1, 2, 3,...$):

$$E_n^{\text{ПТ}}(L) = E_n^{\text{СЧ}}(L) + E_n^{\text{ГО}}(L), \quad E_n^{\text{СЧ}}(L) = W(L)n^2,$$
$$E_n^{\text{ГО}}(L) = \hbar\omega(L)(n+1/2), \quad \hbar\omega(L) = 2W(L)\lambda(L). \qquad (12)$$

---

[1] В соответствии с требованием *строгой положительности* показателя $\lambda(L)$ мы граничиваемся в (11) только одним решением квадратного уравнения для этой величины.

Учитывая определения (9) и (11), можно представить энергетический спектр (5) в более компактном виде:

$$E_n^{\text{ПТ}}(L) = W(L)[n^2 + 2\lambda(L)n + \lambda(L)] = W(L)\{[n+\lambda(L)]^2 - \lambda(L)[\lambda(L)-1]\},$$
откуда $\quad \varepsilon_n(L) = [n+\lambda(L)]^2. \qquad (13)$

Нетрудно показать, что предельные случаи квантовой модели Пёшля–Теллера, рассмотренные в предыдущем разделе, а именно модель и модель гармонического осциллятора Блоха, имеют место не только на уровне разложения потенциала, но и на уровне энергетического спектра $E_n^{\text{ПТ}}(L)$. Действительно, **модель свободной частицы в ящике** соответствует *конечному* значению $L$ и отсутствию потенциала внутри ящика ($V_0 \to 0$), откуда из (11) следует, что $\lambda(L) = 1$, тогда как $W(L)$ вообще не меняется. При этом из (12) формально следует, что $E_n^{\text{СЧ}}(L) = W(L)n^2$ ($n = 0,1,...$), что дает для энергии основного состояния (при $n = 0$) неверное значение $E_0^{\text{СЧ}}(L) = 0$.

Однако, как показано в предыдущем разделе (см. формулу (8) и дальнейший текст), в пределе $V_0 \to 0$ энергией свободной частицы в ящике является вся энергия $E_n^{\text{ПТ}}(L)$, так что указанный выше промежуточный результат должен быть поправлен за счет вклада слагаемого $E_n^{\text{ГО}}(L)$. Действительно, для рассматриваемого случая $\lambda(L) = 1$ получаем из (12), что $\hbar\omega(L) = 2W(L)\lambda(L) = 2W(L)$, откуда сразу следует, что $E_n^{\text{ГО}}(L) = \hbar\omega(L)(n+1/2) = 2W(L)(n+1/2)$. Тогда окончательно находим $E_n^{\text{ПТ}}(L) = W(L)(n^2 + 2n + 1) = W(L)(n+1)^2$ ($n = 0,1,...$), или в более привычном и естественном виде $E_n^{\text{СЧ}}(L) = W(L)n^2$ ($n = 1,2,...$), что приводит теперь к правильному результату для основного состояния $E_1^{\text{СЧ}}(L) = W(L)$.

**Модель осциллятора Блоха** получается несколько более сложным путем; действительно, в пределе $L \to \infty$, $\alpha(L) \to 0$ имеем согласно (4) $W(L) = (\hbar^2/2m)\alpha^2(L) \to 0$, так что $E_n^{\text{СЧ}}(L) = W(L)n^2 \to 0$, т.е вклад в спектр, *квадратичный* по главному квантовому числу $n$ и характерный для свободной частицы в ящике, полностью исчезает. При этом в полном спектре (12) остается только *линейный* по $n$ вклад $E_n^{\text{ГО}}(L)$, характерный для гармонического осциллятора, причем в блоховском пределе (т.е. при $L \to \infty$, $V_0 \to \infty$, см. предыдущий раздел) частота этого осциллятора $\omega(L)$ должна перейти в блоховскую $\omega_{БО} = (k/m)^{1/2}$, где величина $k$ определена в (3). Действительно, для осциллятора Блоха согласно условиям (3) имеем $v(L) = V_0/W(L) \to \infty$, откуда согласно (11) находим $\lambda(L) \approx (1/2) 2v(L)^{1/2} = [V_0/W(L)]^{1/2} \to \infty$, а с учетом определения (3) получаем окончательно ожидаемый результат:

$$\hbar\omega(L) = 2W(L)\lambda(L) \approx 2[V_0 W(L)]^{1/2} = 2(\hbar^2/2m)^{1/2}[V_0 \alpha^2(L)]^{1/2} =$$
$$= 2(\hbar^2/2m)^{1/2}[1/2k]^{1/2} = \hbar(k/m)^{1/2} \equiv \hbar\omega_{БО}. \quad (14)$$

В заключение этого раздела заметим, что было бы интересно более подробно изучить энергетический спектр потенциала ангармонического осциллятора с ангармонизмом четвертого порядка (малым при больших значениях эффективной ширины $L$). Эта модель играет роль промежуточной между двумя предельными случаями свободной частицы и блоховского осциллятора, и мы оставляем соответствующий расчет заинтересованному читателю. Заметим лишь, что влияние ангармонизма существенно изменяет характер энергетического спектра, который перестает быть линейным (и, следовательно, эквидистантным); точнее, к линейному спектру $E_n^{ГО}(L) = \hbar\omega(L)(n+1/2)$ в низшем порядке теории возмущений добавляется слагаемое $\Delta E_n(L) = W(L)(n^2 + n + 1/2)$ (см., например, [2–4]). Правда, в более высоких порядках теории возмущений добавляются «нефизические» слагаемые порядка $n^3$, $n^4$,..., отсутствующие в *точном* выражении (12), что явно свидетельствует об ограниченной применимости теории возмущений применительно к сингулярным потенциалам, к которым относится потенциал модели Пёшля–Теллера.

## Оператор давления в квантовой модели Пёшля–Теллера

Одним из важнейших свойств, присущих ограниченным в пространстве объектам (как классическим, так и квантовым) является возможность введения для них понятия *механического давления P*. В свою очередь, это является необходимым условием для возможности полного описания подобных объектов – как динамического, так и – в особенности – термодинамического, поскольку без понятия давления невозможно определить термическое и бароэнергетическое уравнения состояния; напомним, что для определения калорического уравнения состояния достаточно наличия у объекта понятия *энергии* (как полной, так и внутренней).

Рассматриваемая в данной работе квантовая модель Пёшля–Теллера (ПТ) является одной из немногих моделей, описывающих квантовый объект в условиях *конфайнмента*, т.е. ограничения пространственной области существования объекта конечной эффективной шириной $L$. Эта модель существенно обобщает широко известную (и, по-видимому, простейшую) модель указанного типа – а именно, модель (квази)свободной частицы (СЧ) в ящике шириной $L$ с непроницаемыми стенками; как показано выше, модель СЧ является одним из предельных частных случаев модели ПТ. Заметим, что в другом предельном случае – блоховского осциллятора (БО) с $L \to \infty$ свойство конфайнмента исчезает, а вместе с ним исчезает и возможность определения давления.

Как известно (см., например, [5, §10]), для классических объектов (например, частицы с импульсом $p$ и координатой $x$) динамическое давление $P$ определяется посредством соотношения $P = -\partial H/\partial V$, где $H(p,x;V)$ – функция Гамильтона, включающая два вида энергии: кинетическую $W(p)$ и потенциальную $V(x;V)$, а также зависящий от внешнего параметра $V$ – объема, доступного объекту[2]. В случае низкоразмерных объектов вместо объема $V$ применяется площадь $S$ или длина $L$, а для термодинамического давления функция Гамильтона заменяется на внутреннюю энергию $U$, имеющую смысл среднего значения $H$.

При описании квантовых объектов *функция* Гамильтона $H(p,x;V)$, согласно принципу соответствия переходит в *оператор* Гамильтона $\hat{H}(\hat{p},\hat{x};L) = \hat{W}(\hat{p}) + \hat{V}(\hat{x};L)$, где $\hat{p}$ и      – операторы импульса и координаты квантового объекта (например, частицы массой $m$), $L$ – $c$-численный параметр, характеризующий эффективную ширину области строгой локализации частицы. В координатном представлении       – потенциальная энергия, обладающая свойством конфайнмента – например, представленная выражениями (1), (2), (5) и (8). Существенно, что в общем случае операторы $\hat{W}(\hat{p})$ и       не коммутируют друг с другом, так что полную энергию частицы характеризует спектр $E_n$ собственных значений оператора       , зависящий от параметров $m$, $V_0$ и $L$.

Отвлекаясь от технических деталей, можно сказать, что в *классическом* случае операция вычисления давления                                      достаточно хорошо определена, однако в интересующем нас в этой статье *квантовом случае это уже не так*. Действительно, если согласно принципу соответствия ввести в качестве оператора давления величину $-\partial \hat{H}(\hat{p},\hat{x};L)/\partial L = -\partial \hat{V}(\hat{x};L)/\partial L$, то она, вообще говоря, не будет коммутировать с оператором энергии       . Это означает, что у этих операторов нет общей системы собственных функций и, следовательно, невозможно одновременно измерить «давление» и энергию и получить уравнения состояния – как динамические, так и термодинамические.

В качестве альтернативы обычно используется другое определение, впервые предложенное Гельманом в 1937 году и несколько позднее (1939) независимо Фейнманом; применимость этого определения составляет содержание т.н. *теоремы*

---

[2] Относительно способа введения объема $V$ в качестве управляющего параметра непосредственно в функцию Гамильтона см., например, учебник [6, гл. II, §1] и более подробно (но несколько сложнее) в монографии [7, Гл. 1, § 5], а также в обзорной статье [9].

*Гельмана–Фейнмана* (ее доказательство можно найти в [6, гл. II, § 2г], а также в монографии [8, гл. 1, § 1] и в обзорной статье [9]). Не приводя здесь формального (хотя и довольно простого) доказательства этой теоремы, мы ограничимся его содержательной стороной, которое основано на следующем очевидном факте: *управляющие параметры*, входящие в гамильтониан квантового объекта (как правило, в потенциальную энергию – например, эффективная ширина $L$), входят также и в спектр $E_n(L)$. Тогда для каждого квантового состояния $n$ (для простоты – невырожденного) можно *определить* давление $P_n(L)$ в этом состоянии так:

$$\text{или} \quad \widehat{P}(L) \equiv -\partial/\partial L[\widehat{H}(\widehat{p},\widehat{x};L)]. \tag{15}$$

Существенно, что операторная форма (15) для $\widehat{P}(L)$ имеет физический смысл *только* при действии на собственную функцию $\psi_n(x)$ оператора энергии (слева направо); иными словами, оператор дифференцирования $(-\partial/\partial L)$ действует только *после* действия оператора энергии на волновую функцию $\psi_n(x)$. Заметим, что в предложенном выше «наивном» определении «по принципу соответствия» порядок применения операторов был *обратным*.

Таким образом, значения $P_n(L)$ из (15) не являются собственными значениями оператора                                      , поскольку зависимость $E_n(L)$ от $L$ опреде-
ляется не только явной зависимостью потенциала           от $L$, но прежде всего *гранич-
ными условиями*, согласно которым волновая функция $\psi_n(x;L)$ имеет нули в точках, где
             имеет сингулярности (т.е. обращается в бесконечность); пример подобного расчета приведен нами выше.

Найдем теперь выражение для собственных значений давления (15) для конкретного случая модели Пёшля–Теллера, используя выражения (12) для собственных значений энергии $E_n^{\text{сч}}(L)$ и $E_n^{\text{го}}(L)$. Имеем тогда:

$$\tag{16}$$

$$P_n^{\text{ГО}}(L) = (2/L)E_n^{\text{ГО}}(L)\{1-\mu(L)\}, \quad \mu(L) = 1-[\lambda(L)-1][2\lambda(L)-1]^{-1}. \tag{17}$$

Заметим, что при вычислении давления согласно (15) весьма существенно, обладает ли соответствующее выражение для энергии свойством *однородности* (в смысле Эйлера), поскольку в этом случае связь между давлением и энергией всегда строго *линейна*. Как правило, в качестве однородной встречается функция $f(L)$, имеющая *степенную* зависимость вида $L^s$ с любым вещественным значением $s \neq 0$, так что ее производная имеет вид $-df(L)/dL=(s/L)f(L)$.

В рассматриваемой здесь модели Пёшля–Теллера имеются *две* однородные функции, одна из которых $E_1^{\text{сч}}(L) = W(L)$ с $s = -2$, а другая – $v(L) = V_0/W(L)$ с $s = 2$, однако функция $\lambda(L) = 1/2\{1+[1+4v(L)]^{1/2}\}$ уже однородной не является. Поэтому выражение (16) для $P_n^{СЧ}(L)$ строго пропорционально $E_n^{СЧ}(L)$ с коэффициентом $(s/L)$, тогда как выражение (17) для $P_n^{ГО}(L)$ имеет несколько более сложный вид. Функция $E_0^{ГО}(L) = W(L)\lambda(L)$ уже не является однородной (хотя и содержит однородный множитель $W(L)$), что и отражается в «смешанном» виде $P_n^{ГО}(L)$.

Что касается неоднородного множителя $\lambda(L)$, то его производная имеет вид (11), который здесь удобно записать в ином виде:

$$d\lambda(L)/dL = -(2/L)\lambda(L)\{[\lambda(L)-1][2\lambda(L)-1]^{-1}\}; \qquad (18)$$

видно, что «неоднородность» $\lambda(L)$ всецело обусловлена множителем, стоящим в фигурных скобках, появление которого нетрудно видеть из следующего простого расчета. Дифференцируя исходное выражение (11), связывающее $\lambda(L)$ и $v(L)$, и учитывая, что $dv(L)/dL = -(2/L)v(L)$, имеем:

$$(d/dL)\{\lambda(L)[\lambda(L)-1]\} = dv(L)/dL,$$
$$(d/dL)\lambda(L) = [2\lambda(L)-1]^{-1}(d/dL)v(L), \qquad (19)$$

откуда, учитывая, что $dv(L)/dL = -(2/L)\lambda(L)$, получаем выражение (18) и далее выражение (17) для $\mu(L)$.

Легко видеть, что «неоднородный» множитель $\mu(L)$ находится в пределах отрезка $1/2 \leq \mu(L) \leq 1$, что в частных (предельных) случаях приводит к физически вполне ожидаемым результатам:

1) В *пределе свободной частицы* (при        ) в ящике с конечной шириной $L$, когда $v(L) = V_0/W(L) \to 0$, $\lambda(L) \to 1$, имеем $\mu(L) = 1$ и $P_n^{ГО}(L) = 0$, т.е. «осцилляторная» часть потенциала ПТ-модели не дает вклада в давление; заметим еще раз, что в рассматриваемом пределе связь между давлением и энергией строго линейная.

2) В *пределе гармонического осциллятора Блоха* (при $V_0 \to \infty$ и $L \to \infty$) имеем $v(L) \to \infty$ и $\lambda(L) \to \infty$. Соответственно, $\mu(L) \to 1/2$, однако вклад в давление $P_n^{ГО}(L)$ по-прежнему отсутствует – на этот раз за счет входящего в (17) множителя $(1/L)$.

Заметим, что по той же причине исчезает вклад в давление $P_n^{СЧ}(L)$, причем как за счет множителя $(1/L)$, так и множителя $W(L) \sim 1/L^2$. Тем самым подтверждается исходное положение о том, что в квантовой модели без конфайнмента невозможно корректно определить давление – проще говоря, «без стенок нет давления».

3) Разумеется, представляет интерес рассмотрение *промежуточного случая*, когда имеют место *конечные* значения как для амплитуды потенциала $V_0$ и эффективной ширины $L$; в, при малых значениях $\mu(L)$ выражения (16) и (17) можно (с учетом (12)) представить в наиболее простом и естественном виде $P_n^{\text{ПТ}}(L) = (2/L)E_n^{\text{ПТ}}(L)$. Такое соотношение равносильно *линейному* бароэнергетическому уравнению состояния, связывающему давление и внутреннюю энергию, характерную для *идеального газа* частиц со «смешанным» характером зависимости уровней энергии от главного квантового числа.

## Заключение

В данной статье подробно рассмотрена одна из наиболее интересных моделей в нерелятивистской квантовой механике одной массивной частицы (в одномерном варианте), а именно модель, введенная Г. Пёшлем и Э. Теллером в 1933 году. Эта модель обладает рядом интересных свойств – в частности, ее предельными случаями являются две наиболее известные модели, изучаемые во всех (в том числе начальных) курсах квантовой механики, в том числе и для ряда инженерных и технических специальностей (в том числе по нанотехнологиям). Этими моделями являются, во-первых, сохраняющая конфайнмент модель квазисвободной частицы в «ящике» с непроницаемыми стенками и, во-вторых, модель квантового гармонического осциллятора, введенная Ф. Блохом в 1932 году и не обладающая конфайнментом. В статье предложено элементарное, но детальное рассмотрение взаимосвязей потенциала, волновых функций и энергетического спектра всех указанных моделей. Одним из основных результатов является введение в моделях с конфайнментом оператора давления на основе теоремы Гельмана–Фейнмана и анализ предельных случаев для этой величины. В следующей работе предполагается обобщение полученных результатов на случай отличной от нуля температуры, включая вычисление статистической суммы, термодинамического потенциала и трех уравнений состояния – термического, калорического и бароэнергетического (о котором, к сожалению, часто забывают в стандартном изложении).

## Литература

# About One Interesting and Important Model in Quantum Mechanics
# I. Dynamic Description

Yu.G. Rudoy, E.O. Oladimedji (jr.)

*People's Friendship University of Russia (RUDN),
Moscow, Russian Federation, 117198, Miklukho-Maklay str., 6;
e-mail: rudikar@mail.ru, nockjnr@gmail.com*

In this paper the detailed investigation of one of the most interested models in the non-relativistic quantum mechanics of one massive particle – i.e., introduced by G. Poeschl and E. Teller in 1933 – is presented. This model includes as particular cases two most popular and valuable models: 1) the quasi-free particle in the box with impenetrable hard walls (i.e., the model with confinement) and 2) Bloch quantum harmonic oscillator, which is unconfined in space; both models are frequently and effectively exploited in modern nanotechnology – e.g., in quantum dots and magnetic traps. We give the extensive and elementary exposition of the potentials, wave functions and energetic spectra of all these interconnected models. Moreover, the pressure operator is defined following the lines of G. Helmann and R. Feynman which were the first who introduced this idea in the late 30-ies in quantum chemistry. By these means the baroenergetic equation of state is obtained and analyzed for all three models; in particular, it is shown the absence of the pressure for the Bloch oscillator due to the infinite width of the "box". The generalization of these results on the case of nonzero temperature will be given later.

*Keywords*: Bloch quantum harmonic oscillator, anharmonic quantum oscillator, Poeschl–Teller quantum oscillator, pressure operator, Hellmann–Feynman theorem.